# CRYSTAL GROWTH AND CHARACTERIZATION OF DOPED CZT CRYSTALS

N. N. Kolesnikov[*], S. P. Vesnovskii[♣], R. B. James[♦], N. S. Berzigiarova[*], D. L. Alov[*], A. G. Zvenigorodskii[♣]

[*]Institute of Solid State Physics, Russian Academy of Sciences, Chernogolovka, Moscow district, 142432 Russia.
[♣]VNIIEF-Conversion J.S. Company, Prospect Mira 46, Sarov, Nizhni Novgorod Region, 607190 Russia.
[♦] Brookhaven National Laboratory, Upton, NY

## Abstract

$Cd_{1-x}Zn_xTe$ crystals with x in the range of 0.1-0.2 were grown by the high-pressure vertical Bridgman method from pre-synthesized CZT. Resistive graphite heaters were used to control the temperature profiles within the furnaces, and an argon overpressure was used to reduce the cadmium loss. The crystals were doped with either Al, Ni, In, Ga, Ge or Sn. The doping was carried out by three different ways: 1) by adding of the pure metals during growth runs; 2) by adding of the tellurides of the metals during growth runs; or 3) by inserting of the metal tellurides during synthesis of the starting CZT material. Some of the growth process parameters were also varied. The as-grown CZT ingots had diameters of either 15 or 38 mm.

The influence of the doping on CZT properties, particularly the conductivity type and specific electrical resistivity, will be discussed. Energy spectra from alpha particles (U-233, Ra-226, and U-233+Pu-239+Pu-238) and from different gamma sources (Cs-137, Co-60, Co-57, Am-241) will be reported.

## 1. Introduction

$Cd_{1-x}Zn_xTe$ is a promising material for the room-temperature radiation detectors [1]. Growth of detector grade CZT is usually performed by the high-pressure vertical Bridgman method (HPVB) [2]. The present paper describes the influence of different dopants (Al, Ni, In, Ga, Ge, and Sn) on some of the properties of HPVB-grown $Cd_{1-x}Zn_xTe$ (x = 0.1 and 0.2). The results are compared with undoped CZT samples.

## 2. Experimental

The initial CZT loads were prepared by solid-state reaction from the elements with subsequent melting of the synthesized material. The compounding was carried out in evacuated silica containers. Two sets of the raw materials were used – elements with 5N+ purity and (for 1 experiment) elements with 6N+ purity.

$Cd_{1-x}Zn_xTe$ crystals with diameters of 15 and 38 mm (figs. 1, 2) were grown in one- and two-zone HPVB furnaces (fig. 3) with resistive graphite heaters. Graphite and silica crucibles were used. The crucible pulling rates were approximately 1-2 mm/h, and the thermal gradients in the solidification zone were in the range of 35 – 70 deg/cm. In some growth runs a steady-state rotation of 10 rpm was applied. The argon pressure was kept at 10.0±0.5 MPa in all growth experiments.

The following techniques of doping were implemented:
(1) Doping with pure elements. The elements are added to the initial loads before growth runs;
(2) Doping with presynthesized $Al_2Te_3$, NiTe, InTe, GaTe, GeTe and SnTe. The tellurides are added to the initial loads before growth runs;
(3) Doping with presynthesized $Al_2Te_3$, NiTe, InTe, GaTe, GeTe and SnTe. These tellurides are added to Cd, Zn and Te mixtures before the initial synthesis of the load.

The concentrations of the dopants varied from 30 to 200 ppm, namely 30 and 90 ppm for the Al-doped runs, 100 ppm for the Ni, In, Ga and Ge runs, and 100 and 200 ppm for Sn doping experiment.

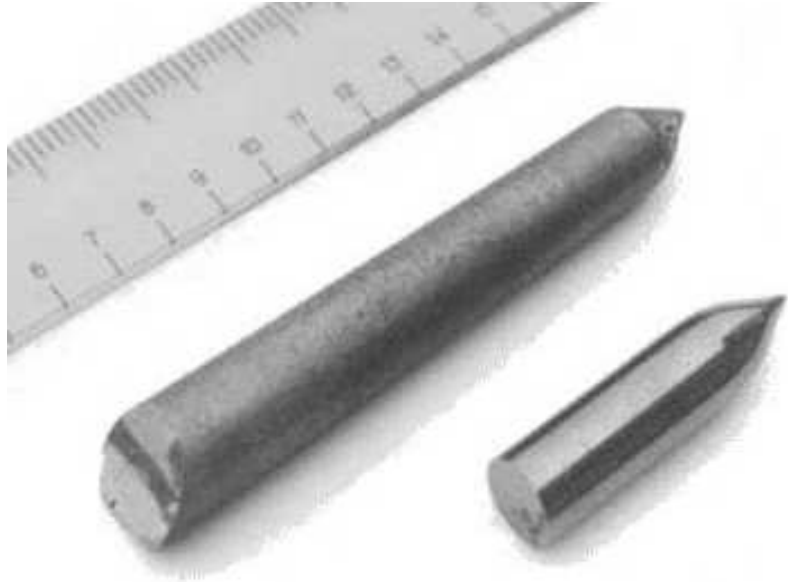

Fig. 1. CZT ingots with diameter of 15 mm.

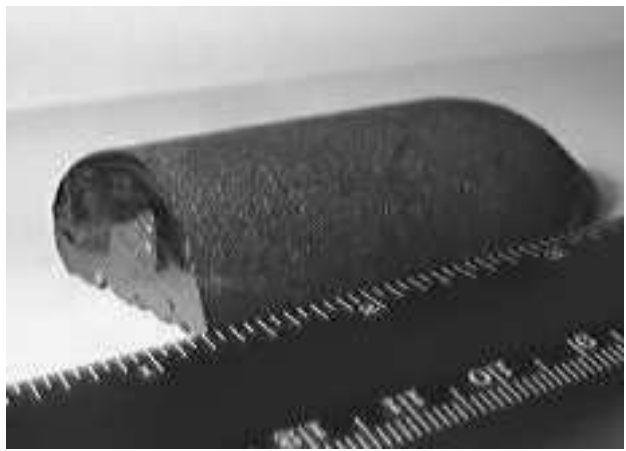

Fig. 2. Half of cylindrical CZT crystal with a diameter of 38 mm. The ruler is in mm units in the foreground and in inches on the opposite side.

Undoped CZT crystals were also grown for comparison.

CZT specimens were prepared from as-grown ingots by mechanical cutting, grinding and polishing, followed by chemical polishing in Br-methanol solutions and subsequent rinsing in pure methanol. All samples had sizes of 10×10×2 mm. In cases where preparation of single crystal samples was possible, the plates were cut parallel to the (110) plane.

The conductivity type (for samples with low and medium specific resistivity only) was determined by the hot-probe method. The specific resistivity ($\rho$) values were measured in dark conditions using a DC four-probe technique. For these measurements indium contacts were applied by melting indium on the surfaces of the CZT plates. The indium contacts were removed from the high-resistivity samples ($\rho \geq 10^9$ Ohm×cm) using a liquid In-Ga mixture, followed by mechanical and chemical repolishing. Next, Al and Au contacts were thermally deposited in high vacuum (fig. 4).

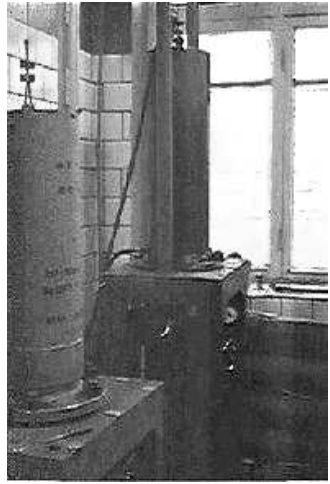

Fig. 3. Two HPVB devices in the growth room.

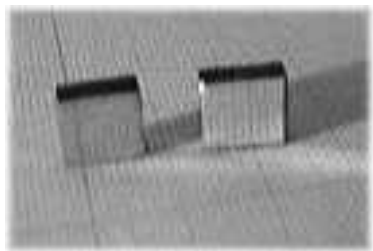

Fig. 4. CZT detector plates. The surfaces were coated with either Au (left) or Al (right).

Energy spectra from alpha particles (U-233, Ra-226, and U-233+Pu-239+Pu-238) and from different gamma sources (Cs-137, Co-60, Co-57, Am-241) were taken at various detector temperatures and bias voltages. The typical scheme of the nuclear response measurements is shown in figs. 5, 6.

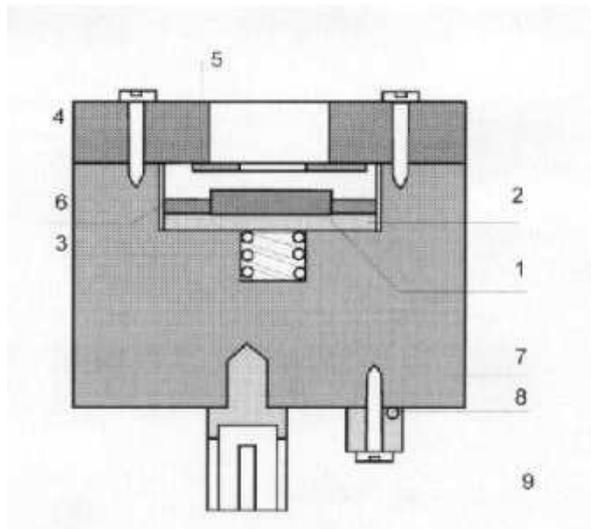

Fig. 5. Sample mount in assembly. 1 – CZT sample; 2 - mobile contact washer; 3 - spring; 4 - insulating washer; 5 - immobile contact washer; 6 - insulating washer; 7 - Al body; 8 - thermocouple; 9 - cooling rod.

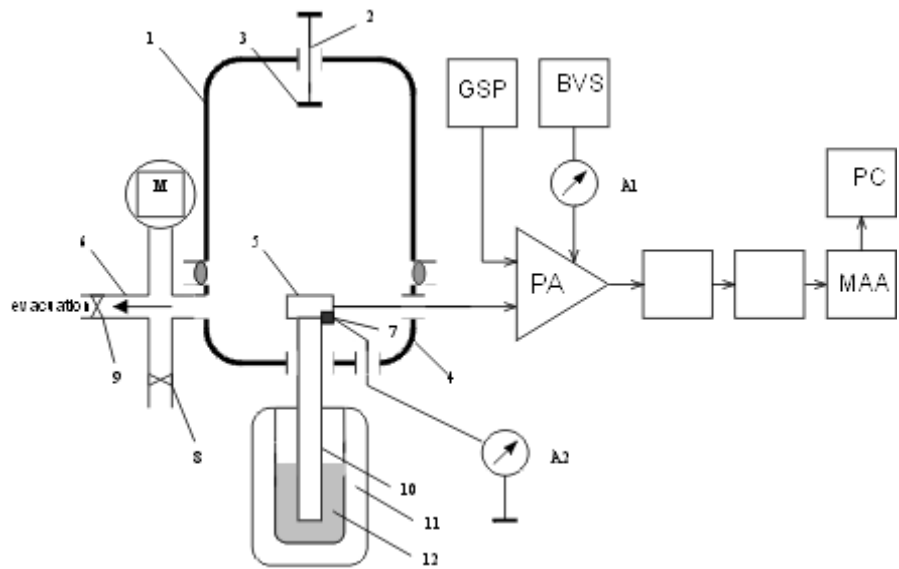

Fig. 6. Geometry and flow chart of the nuclear response measurements. 1 – vacuum chamber cap, 2 – guide-rod of the source displacement, 3 – alpha- and gamma-source, 4 – vacuum chamber pallet, 5 – sample under investigation, 6 – nozzle for vacuum chamber pump-out, 7 – thermocouple, 8 – valve-leak, 9 – pump-out valve, 10 – copper rod, 11 – Dewar vessel, 12 – liquid nitrogen, PA – pre-amplifier, GSP – generator of standard pulses, BVS – bias voltage source, MAA – multi-channel amplitude analyzer, PC – personal computer, A1, A2 – micro-ammeters.

### 3. Results and Discussion

The $Cd_{1-x}Zn_xTe$ (x=0.1-0.2) crystals, which were grown from the initial loads of synthesized 5N+ purity raw materials, usually do not detect gamma radiation and, in the best case, might be useful only as an alpha counter [4]. Since some of the crystals had specific resistivity of $10^9$-$10^{10}$ Ohm×cm, the problem was more than noise due to excessive leakage current.

The $Cd_{0.9}Zn_{0.1}Te$ ingot prepared from 6N+ purity materials had detector grade properties– see the energy spectra shown in figs. 7-10.

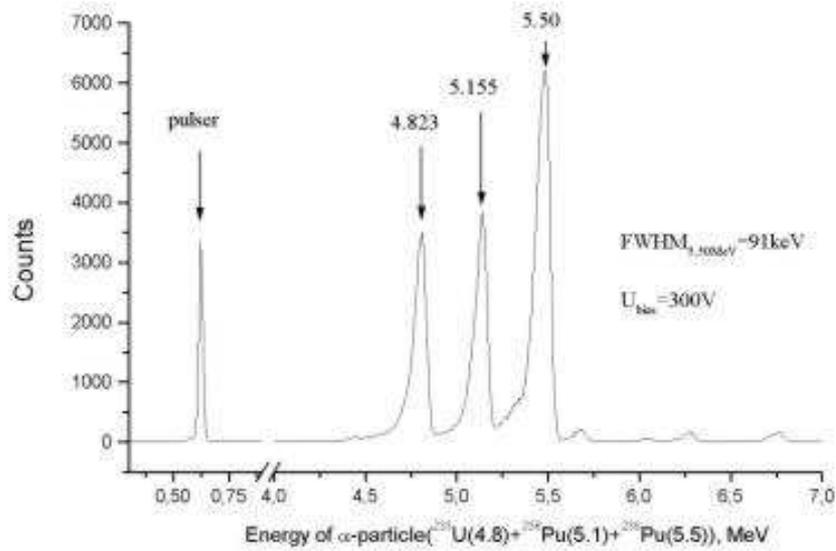

Fig. 7. Alpha particle energy spectra of undoped $Cd_{0.9}Zn_{0.1}Te$.

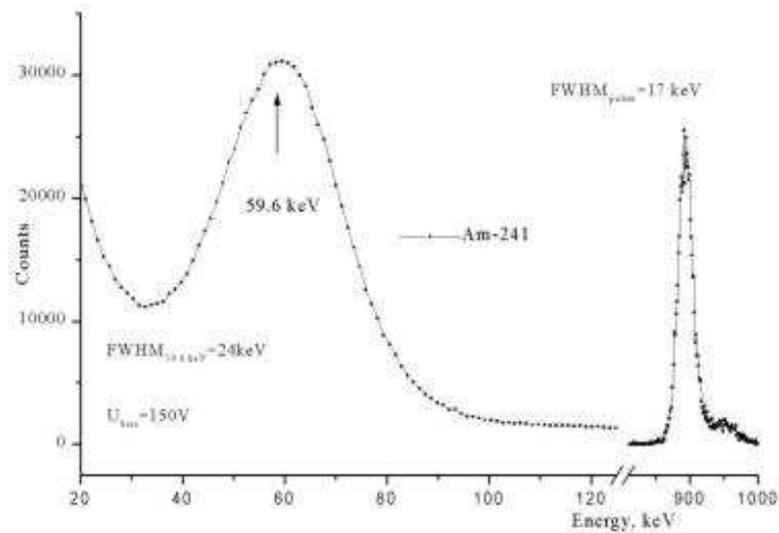

Fig. 8. Gamma (Am-241) energy spectrum of undoped $Cd_{0.9}Zn_{0.1}Te$.

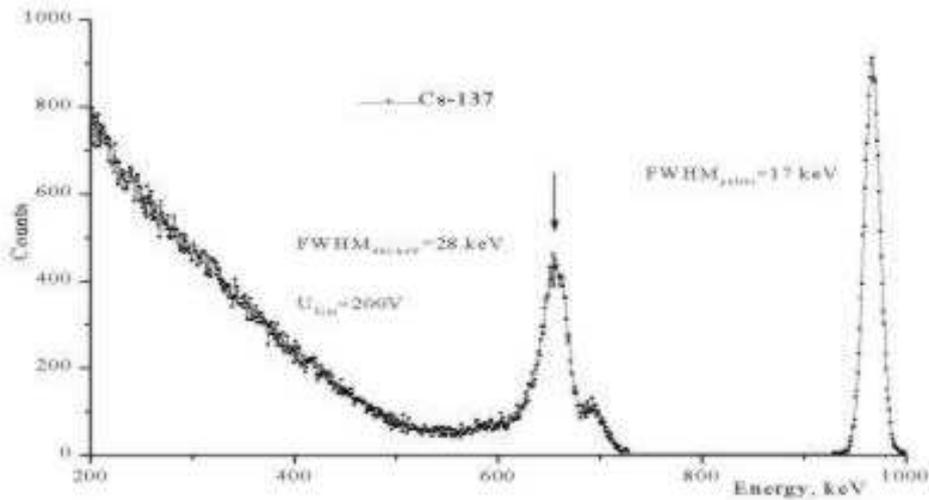

Fig. 9. Gamma (Cs-137) energy spectrum of undoped $Cd_{0.9}Zn_{0.1}Te$.

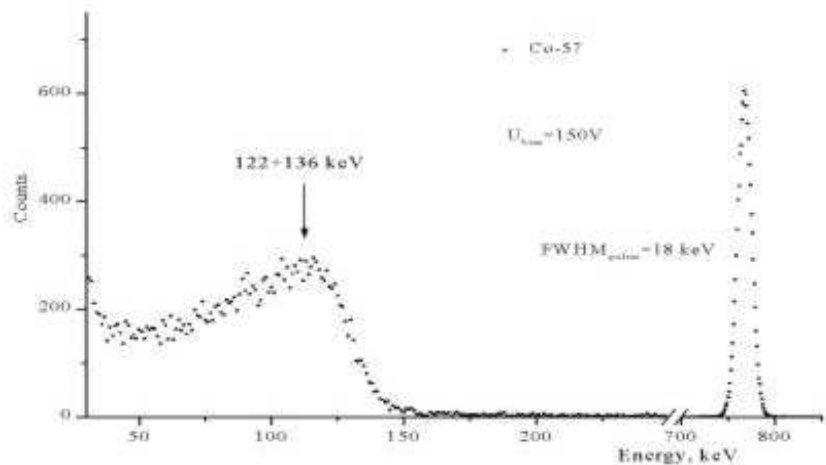

Fig. 10. Gamma (Co-57) energy spectrum of undoped $Cd_{0.9}Zn_{0.1}Te$.

The CZT crystals, which were grown by HPVB from stoichiometric initial loads, tend to have *p*-type conductivity due to shifting of the main composition toward an excess of tellurium. A "self-doping" by cadmium vacancies likely results from the excess tellurium [3]. Attempts to compensate these crystals by "self-doping" with donors, i.e., by adding Zn or Cd or Zn+Cd doping [1-5] during crystal growth or by annealing of CZT samples in Zn or Cd vapours resulted in CZT crystals with specific resistivity up to $5 \times 10^{10}$ Ohm×cm. However, the samples showed a relatively low electron mobility and lifetime. Detectors produced from these CZT crystals did not detect gamma radiation.

In this work we continued attempts to obtain compensated CZT prepared from 5N+ purity raw materials by adding three different donor dopants – Al, In and Ga. The background content of these elements in undoped $Cd_{1-x}Zn_xTe$ (x=0.1-0.2) was found to be ≈ 150 ppm for Al, less than 1 ppm for Ga, and less than 5 ppm for In. We also attempted to study the influence of isovalent Ni, Ge and Sn on the trapping of electrons and holes. These dopants are expected to create deep levels in the energy gap

[6], which would be expected to have a negative effect on the CZT nuclear response. The effects on electrical resistivity were expected to be a strong function of the energy level for the dopant and its concentration.

The donor doping (with Al at 90 ppm, and In or Ga at 100 ppm) by methods described earlier led to CZT polycrystals with narrow "compensated" stripes near the middle of the ingots. The shape of these stripes roughly followed the shape of the crystallization front. The stripes were located at about 30 % of the crystal length, counting from the bottom of the boule (i.e., where the crystallization starts). They had widths along the growth direction of 3-5 mm for ∅ 15 mm ingots and 10-15 mm for ∅ 38 mm ones. The specific electrical resistivities of the CZT samples cut from these stripes were between $10^9$ and $10^{10}$ Ohm×cm. The portions of the ingot below and above these stripes had lower ρ values, and they were of *n*-type conductivity. The ingot inverted to *p*-type near the last-to-freeze section (approximately the last 10 % of the ingot length), where the excess of Te is much more substantial due to the segregation and cadmium losses during growth. The values of ρ in the *n*-type region lowered with increasing distance from the "compensated" stripe (down to $10^3$ Ohm×cm). It seemed that changing of the thermal gradient (G) within the aforementioned interval did not appreciably affect the electrical properties of the crystals. Adding of a second HPVB pass (with ingots from one-pass growth as the initial load) influenced only the "compensated" stripe position, shifting it further toward the middle of the crystal by another 10-15 % of the boule length. The crystal quality improved after the second pass – the volume of the single crystal blocks (grains) increased from ≤ 1-2 $cm^3$ (after one-pass growth) up to 3-5 $cm^3$. The grain size after the two-pass growth also increased by increasing the thermal gradient, but increasing G resulted in the formation of more polysynthetical twins within the single crystal blocks. Adding of steady-state rotation at 10 rpm changed the position of the "compensated" stripe, making it asymmetrically positioned, i.e., shifted the stripe to one side of the cylindrical crystal and slightly elongated it along the growth direction.

The decreasing of the aluminum donor dopant concentration down to 30 ppm led to the disappearance of the "compensated" stripe, replacing it with an *n*-type region (ρ ≈ $10^6$ Ohm×cm).

The use of method (3) for donor doping, as described above, gave initial loads that had a more homogeneously distributed dopant. This doping method led to ingots with "compensated" stripes shifted toward the upper portions of the boules.

The nuclear response of the donor-doped samples with ρ ≥ $10^9$ Ohm×cm was inadequate for their use as detectors of ionizing radiation. An example of the energy spectra ($Cd_{0.8}Zn_{0.2}Te$ with 90 ppm of Al) is shown in fig. 11.

Doping with 100 ppm of Ni, as it was expected, had no registered effect on the measured electrical properties of the CZT crystals, independent of the method of doping.

The Ge and Sn doping produced results that are not explained for now. The addition of 100 ppm of Ge and Sn, which were added by methods (1), (2) and (3) in one-pass HPVB growth, resulted in CZT crystals with properties resembling the ones of the donor doped CZT described above. The only one difference was that maximum ρ values, achieved in the "compensated" stripes of the Ge-doped ingots, was only $10^8$ Ohm×cm. Based on the nuclear response testing, the Ge- and Sn-doped crystals also proved to not be of high detector quality.

Basing on the obtained data we had chosen the 200 ppm way (method 3) of Sn-doping for the two-pass HPVB experiment. Polysynthetically twinned (with rotation twins around <111> axis) $Cd_{0.9}Zn_{0.1}Te$ crystals were obtained for an ingot with a diameter of 38 mm. The specific resistivity of this ingot varied from $2×10^9$ Ohm×cm near the beginning and the end of the ingot, up to $5×10^{11}$ Ohm×cm in the middle part. Nuclear response measurements of the samples cut from this ingot have not yet been completed.

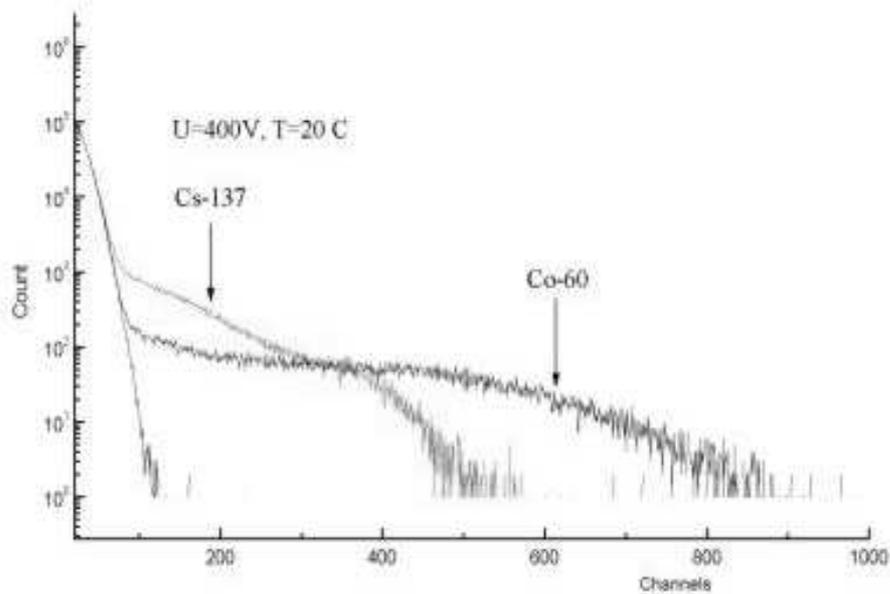

Fig 11. Gamma (Cs-137 and Co-60) energy spectra of $Cd_{0.8}Zn_{0.2}Te:Al$.

## 4. Conclusion.

The influence of dilute ($\leq 100$ ppm) doping of several elements on the electrical properties of CZT crystals and the nuclear response of fabricated detectors is not significant in comparison to the role of foreign impurities. It is impossible to improve $Cd_{1-x}Zn_xTe$ (x=0.1-0.2) crystals, prepared from 5N+ purity materials, from poor counters to detector grade quality by doping with specific elements at the desired concentrations. At the same time increasing the purity of the raw materials up to 6N+ or more allows growth of detector grade CZT without the intentional introduction of dopants.

The two-pass HPVB process yields CZT crystals with better crystal quality than those produced with a one-pass growth technique.

Doping of Sn at 200 ppm Sn by addition of SnTe to the initial synthesized load yields high-resistivity twinned single crystals after two-pass HPVB. Investigations of material and detector properties have not yet been completed.

## Acknowledgements

This work was supported by CRDF Grant Assistant Program grants RE0-11112-SNL and RE0-0182-SNL. The authors thank Jim Rea of Sandia National Laboratories, New Mexico for his consultation and guidance. One of the authors (R. B. James) acknowledges partial support from Sandia National Laboratories under the DOE Initiatives for Proliferation Prevention Program.